\documentclass[aps,prl,twocolumn,amsfonts,showpacs,superscriptaddress,floatfix,url]{revtex4-1}
\usepackage{graphicx}
\usepackage{subfigure}
\usepackage{natbib}
\usepackage{mathrsfs}
\usepackage{amsmath,amssymb,amsfonts}
\usepackage{bm}
\usepackage{braket}
\usepackage{color}
\usepackage{url} 
\usepackage{epsfig,amsopn}

\usepackage{amsmath}

\DeclareMathOperator{\arctanh}{arctanh}

\newcommand{ \td } {\, \mathrm{d} }

\begin{document}

\title{ Stability of topological wall defects on spheres with $ n $-atic order }
\author{ C. Saichand }
\affiliation{ Raman Research Institute, C.V. Raman Avenue, Bangalore 560 080, India }
\author{ Jaya Kumar Alageshan }
\affiliation{ Indian Institute of Science, C.V. Raman Avenue, Bangalore 560 012, India }
\author{ Arun Roy }
\affiliation{ Raman Research Institute, C.V. Raman Avenue, Bangalore 560 080, India }
\author{ Yashodhan Hatwalne} 
\email{yhat@rri.res.in}
\affiliation{ Raman Research Institute, C.V. Raman Avenue, Bangalore 560 080, India }

\date{today}

\begin{abstract}
Topological point defects on orientationally ordered spheres, and on deformable fluid vesicles have been partly motivated by their potential applications in creating super-atoms with directional bonds through functionalization of the ``bald-spots" created by topological point defects, thus paving the way for  atomic chemistry at micron scales. We show that singular wall defects, topologically unstable ``bald lines" in two dimensions, are stabilized near the order-disorder transition on a sphere. We attribute their stability to free-energetic considerations, which override those of topological stability. 

\end{abstract}

\date{\today}

\maketitle

The remarkable interplay between the curvature of a surface, and frustration of orientational order on it is strikingly demonstrated by the Poincar\'e-Hopf index theorem  \textcolor{blue}{\cite{HairyBall}}. Stated informally, a hairy ball cannot be combed flat without creating at least one hair-whorl; a singular, isolated disclination (vortex),  or isolated disclinations with total index (winding number) 2. For vector (1-atic) order disclinations have integer indices, whereas for nematic (2-atic) order they are integer multiples of 1/2.  In a region surrounding a disclination (point defect in 2-dimensions), deformations in the orientation-field are large enough to destroy orientational order. Disclinations are topological defects characterized by their index, and have ``molten" core regions of finite extent encompassing the disclination points.

 The study of point disclinations on ordered spheres  \textcolor{blue}{\cite{LubenskyProst}} and on deformable vesicles \textcolor{blue}{\cite{ParkLubMcK}} is important for investigating the interplay between geometry, topology, and elasticity, and for its potential applications in materials science. It gained impetus from the proposal that disclination cores on spherical particles such as micron-scale colloidal particles coated with liquid crystals can be functionalized to create ``super-atoms"  with directional bonds \textcolor{blue}{\cite{NelsonNano}}. This opened up new possibilities such as self-assembly of super-atoms by linking across functionalized groups (including biomolecules such as DNA), and the development of atomic chemistry at micron scales. Rigid spheres have been prepared by molecular coating of ordered, tilted monolayer on metal nanospheres \textcolor{blue}{\cite{DeVries}}, leading to the  antipodal configuration of a source-sink pair of disclinations of index 1 each. These divalent super-atoms spontaneously  form long one-dimensional chains.  Thin nematic shells consisting of a nematic drop containing a smaller aqueous drop have been obtained in double emulsions  \textcolor{blue}{\cite{Lopez-Leon}}. These can be engineered to imitate $ sp $-, $ sp^{2} $-, and $ sp^{3} $ geometries of carbon bonds. Deformable vesicles with orientational order can form facets. These fascinating possibilities have led to rapid advances in theoretical and experimental studies  \textcolor{blue}{\cite{VitelliNelson,ShinBowickXing,XingEtAl,HerstEtAl,ManyuhinaBowick}}. 

In this letter we address the energetics and stability of \textit{topological wall defects} (line defects in 2-dimensions) on spherical fluid membranes with $ n $-atic orientational order. Singular wall defects in two- and three dimensional ordered systems are topologically unstable because they can be made to disappear by making local changes in the orientational order \textcolor{blue}{\cite{deGennesProst,ChaikinLubensky}}. In three dimensions, removal of disclinations lines with index 1 via ``escape" of the nematic director in the third dimension \textcolor{blue}{\cite{deGennesProst,ChaikinLubensky, KlemanLavrentovich,CladisKleman,Meyer}} is well known. However, close to nematic-smectic transition the bend elastic constant diverges, the escape configuration has a larger free energy than that of the line disclination of index 1, and the disclination line is stabilized.  

We show that singular wall defects can be stabilized on a sphere because of its Gaussian (intrinsic) curvature, and not because of  boundary conditions, externally imposed fields, or divergences in certain elastic constants. They are stable close to the order-disorder transition, over a finite range of a dimensionless parameter $ \eta $. The parameter $ \eta $ is the ratio of basic free energy scales corresponding to the destruction of order in the defect cores to that of the elastic deformation outside the core. Our results on vector-, and nematic orders are summarized in Figs.1-5. The extension of these results to $ n $-atic order is straightforward. Remarkably, we find that for $ n $-atic order, the lowest \textit{elastic free-energy} configuration has $ 2 n $ walls of index $ 1/n $ each, located such that the integral of Gaussian curvature of the sphere between any two successive walls is $ 2 \pi /n $. The one-dimensional, closed loop walls that we consider have an unusual feature -- each loop is characterized by a continuous disclination density along its length, with a quantized index assigned to the entire loop \textcolor{blue}{\cite{KlemanFriedel}}. To the best of our knowledge, such defects have not been discussed in condensed matter systems.
 
\begin{figure}[t]
\includegraphics[width=2.3in]{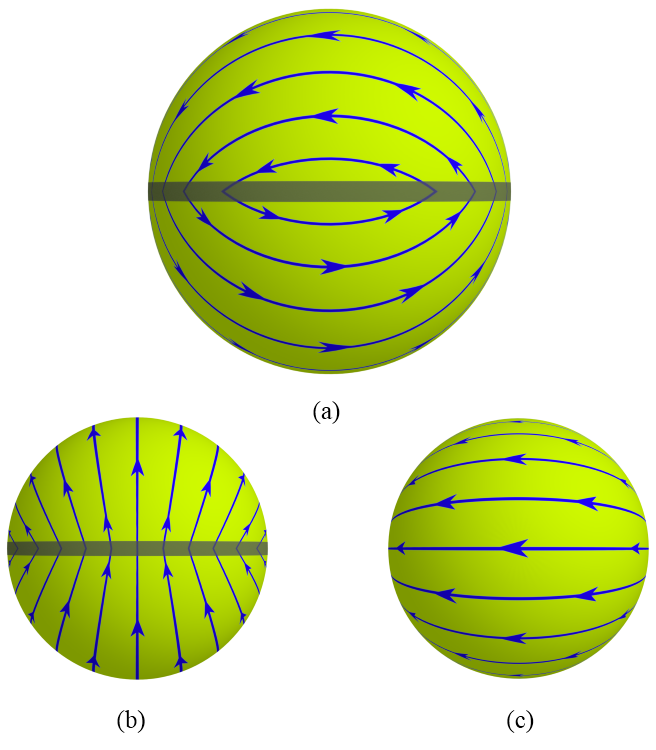} 
\caption{(Color online) Equatorial wall: (a) Side ($ \theta = \pi/2, \, \phi = \pi/2 $) view of the index 2, equatorial wall. The directed lines are the streamlines of the vector field. The shaded region represents the disordered core, within which vector order is completely destroyed, and cannot be assigned a direction. The full field, including that shown within the core region, corresponds to a wall with zero core-size. Rounding off the slope singularity of the field at the equator, the wall defect of zero core-size transforms into the antipodal configuration of a pair of index 1 point disclinations. (b) Front view ($ \theta = \pi/2, \, \phi = 0 $). (c) Top view, showing that the polar regions are free of point disclinations.}
\label{ew}
\end{figure}

 \begin{figure}[t]
\includegraphics[width=2.3in]{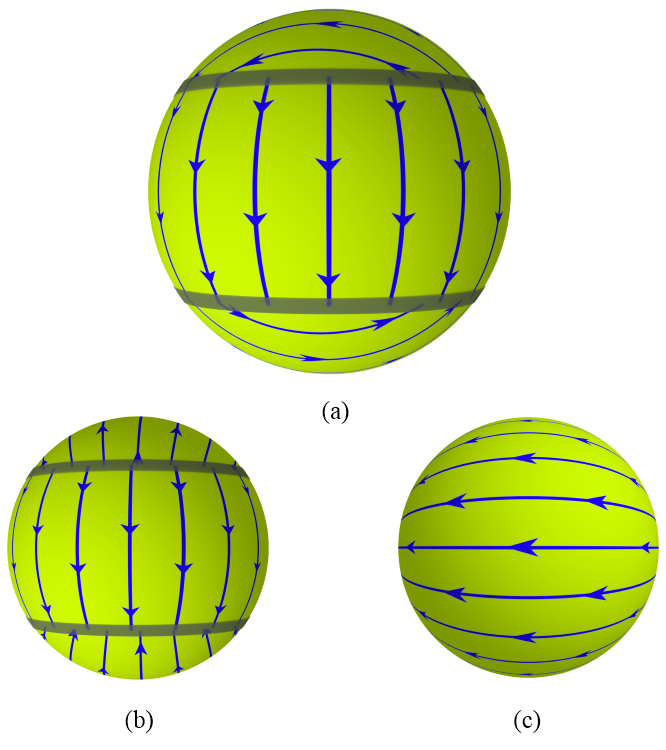} 
\caption{(Color online) Two-wall configuration: (a) Side view. The vector field between the two index 1 walls follows the longitudes. The two-wall configuration degenerates to the antipodal configuration by smoothly sliding the walls towards the respective poles. (b) Front view. (c) Top view.}
\label{2w}
\end{figure}

On curved, orientationally ordered surfaces the free energy takes the form of Ginzburg-Landau theory for superconductivity \textcolor{blue}{\cite{ParkLubMcK}}. Apart from the metric tensor, the complex order parameter is coupled to the shape through a gauge field arising from the covariant derivative of the order parameter.  For $ n $-atic order a disclination of index $ (p/n) $ where $ p $ is an integer, is a singular point around which the phase angle of the order parameter changes by 
$ 2 \pi (p/n) $. In superconductors the total vorticity is determined by an external magnetic field, whereas for a surface the signed, total index of the disclinations is fixed by its topology. The free energy density of orientationally ordered surfaces can be cast as  
\begin{equation}
\label{fGL}
f_{\mathrm{GL}} = \frac{r}{2} \, \vert \psi \vert^{2} + \frac{u}{4} \, \vert \psi \vert^{4} 
                                                                                + \frac{k}{2} \, \vert (\bm{\partial} - i \,  \bm{A}) \, \psi \vert^{2},                             
\end{equation}
where $ r = r_{0} (T - T_{c}) $,  $ T $ and $ T_{c} $ 
are the temperature, and the critical temperature respectively, $ u $ is assumed to be temperature independent, and
$ \psi  = \psi (\underline{ \bm{\sigma} }) = 
\vert \psi(\underline{ \bm{\sigma} }) \vert \, e^{i \alpha(\underline{ \bm{\sigma} })} $ 
is the complex order parameter. For $ n $-atic order the angle $ \alpha $ is measured modulo $ 2 \pi/n $ with respect to a local orthonormal frame 
$ (\hat{ \bm{e} }_{1}  (\underline{ \bm{ \sigma} }), \hat{ \bm{e} } _{2}  (\underline{ \bm{\sigma} }))  $ 
on the surface, where 
$ \underline{\bm{\sigma}} = (\sigma^{1}, \sigma^{2}) $ represent the internal coordinates parameterizing the surface. 
The term with coefficient $ k $ describes the elastic free energy of deformations within the ``one-constant" approximation \textcolor{blue}{\cite{deGennesProst,ChaikinLubensky,KlemanLavrentovich}}.  In this term $ \bm{\partial} $ represents the flat-space gradient operator, and $ \bm{A}( \underline{\bm{\sigma}}) $ is the local gauge field (the spin connection) that corrects the flat-space gradient by accounting for membrane curvature. The components of the spin connection are given by 
$ A_{\mu}(\underline{\bm{\sigma}}) = \hat{\bm{e}}_{1} (\underline{\bm{\sigma}}) \cdot  \partial_{\mu} \hat{\bm{e}}_{2}(\underline{\bm{\sigma}}) $, where $  \mu \in (\sigma^{1}, \sigma^{2}) $.
In the term with coefficient $ k $, complex conjugates are contracted using the metric tensor.
The free energy $ F_\mathrm{GL} = \int f_\mathrm{GL} \td \mathcal{A} $, where the area element $ \td \mathcal{A} = \sqrt{g}  \td \sigma^{1} \td \sigma^{2} $, and $ g $ is the determinant of the metric tensor $ g_{\mu \nu} $. The condensation free energy density (the free energy cost per unit area for destruction of orientational order) $ f_{\mathrm{C}}  = r^{2}/(4 u) $.  It plays an important role in determining the core-size of topological defects  \textcolor{blue}{\cite{ChaikinLubensky}}. The elastic part of the free energy density \textcolor{blue}{(\ref{fGL})} is  
\begin{equation}
\label{fel}
f_{\mathrm{el}} =  \frac{1}{2} K_{\alpha} \, (\bm{\partial} \alpha - \bm{A})^{2} =  \frac{1}{2} K_{\alpha} \, (\partial_{\mu} \alpha - A_{\mu})(\partial^{\mu} \alpha - A^{\mu}), 
\end{equation}
where $ K_{\alpha} = k \, \vert \psi \vert^{2} $ in the mean field approximation.  
Minimization of the elastic free energy $ F_{\mathrm{el}} = \int f_{\mathrm{el}} \td \mathcal{A} $ leads to the Euler-Lagrange equation of equilibrium
$ (\delta F_{\mathrm{el}}/\delta \alpha) =  - K_{\alpha} \, \bm{\nabla} \cdot  (\bm{\partial} \alpha - \bm{A}) = 0, $                         
where the operator $ \bm{\nabla} \cdot $ denotes the covariant divergence. For flat surfaces the gauge field $ \bm{A} = 0 $, and $ F_{\mathrm{el}} $ reduces to the elastic free energy density of the continuum $ xy $-model.
The Airy stress function 
$ \chi $  defined by $ \partial^{\mu} \alpha - A^{\mu} = \gamma^{\mu \nu} \partial_{\nu} \chi $, 
with the unit antisymmetric tensor density $ \gamma^{\mu \nu} = \epsilon^{\mu \nu}/\sqrt{g} $, identically satisfies $ (\delta F_{ \mathrm{el} } /\delta \alpha) = 0 $.  However, $ \chi $ has to obey the compatibility condition 
\textcolor{blue}{\cite{DeemNelson,ParkLubensky,BowickGiomi}}
\begin{equation}
\label{CompatibilityCond}
\nabla^{2} \chi(\underline{\bm{\sigma}}) = K(\underline{\bm{\sigma}}) - \mathscr{S}(\underline{\bm{\sigma}}).
\end{equation}
Here, $ \nabla^{2} $ is the covariant Laplacian operator, $ K(\underline{\bm{\sigma}}) $ is the Gaussian curvature, and 
$ \mathscr{S}(\underline{\bm{\sigma}}) $ is the disclination density  \textcolor{blue}{\cite{Footnote1}}.  In terms of the stress function $ \chi $, the elastic free energy density \textcolor{blue}{(\ref{fel})} can be written as 
\begin{equation}
\label{felChi} 
f_{\mathrm{el}} =   \frac{1}{2} K_{\alpha} \, (\bm{\partial} \chi)^{2}  =  \frac{1}{2} K_{\alpha} \, (\partial_{\mu} \chi)(\partial^{\mu} \chi).
\end{equation}

With this background we discuss the procedure followed in calculating the free energies of point- and wall defects on a sphere of radius $ R $. For vector order the free energy of the antipodal configuration of (index 1) point disclinations has been obtained for small core-sizes $ r_{\mathrm{c}} $, where the dimensionless cutoff $ \zeta = r_{\mathrm{c}}/R \ll  1 $  \textcolor{blue}{\cite{LubenskyProst}}.
A larger core reduces the elastic free energy $ F_{\mathrm{el}} $, but increases the condensation free energy. Making the simplifying assumption that order is destroyed over the entire core region, the condensation free energy 
$ F_{\mathrm{C}} = \int f_{\mathrm{C}} \td \mathcal{A} $, where the integral is over the core region \textcolor{blue}{\cite{ChaikinLubensky}}.  \textit{In investigating the energetics and stability of point- as well as wall defects for $ n $-atic order  the determination of optimal core-sizes, without the restriction $ \zeta \ll 1 $,  is  crucial}.  In what follows, we minimize the dimensionless total free energies
 $ F_{\mathrm{T}} =  (F_{\mathrm{el}} +  F_{\mathrm{C}})/K_{\alpha} $ with respect to $ \zeta $   to obtain the optimal core size.  We focus primarily on vector-, and nematic orders, and discuss the antipodal-, equatorial wall-, two-wall-, tetrahedral-, and four-wall configurations. We find that these configurations are ground states in different ranges of $ \eta = f_{\mathrm{C}} R^{2}/K_{\alpha} $. To compare their total free energies and ranges of stability, it is convenient to choose the dimensionless condensation energy $ 4 \pi \eta $ corresponding to the destruction of order over the entire sphere as the common reference of free energy. 
 
 \begin{figure}[t]
\includegraphics[width=2.3in]{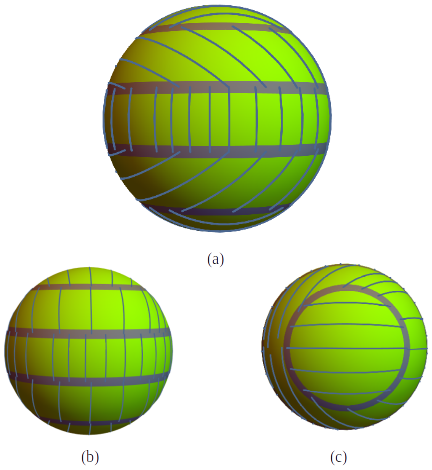} 
\caption{(Color online) Four-wall configuration (for nematic order): (a) Side view. (b) Back ($ \theta \simeq \pi/2 $, $ \phi \simeq 3 \pi/2 $) view. (c) Top view. \textit{Each wall in this configuration has the index 1/2, with uniform, linear disclination density. Therefore the change in the angle per unit length across each wall is small as compared to that for index 1 (and 2) walls}.  This should be borne in mind, particularly while viewing the back view (b). }
\label{4w}
\end{figure}

\begin{figure}[t]
\includegraphics[width=3.3in]{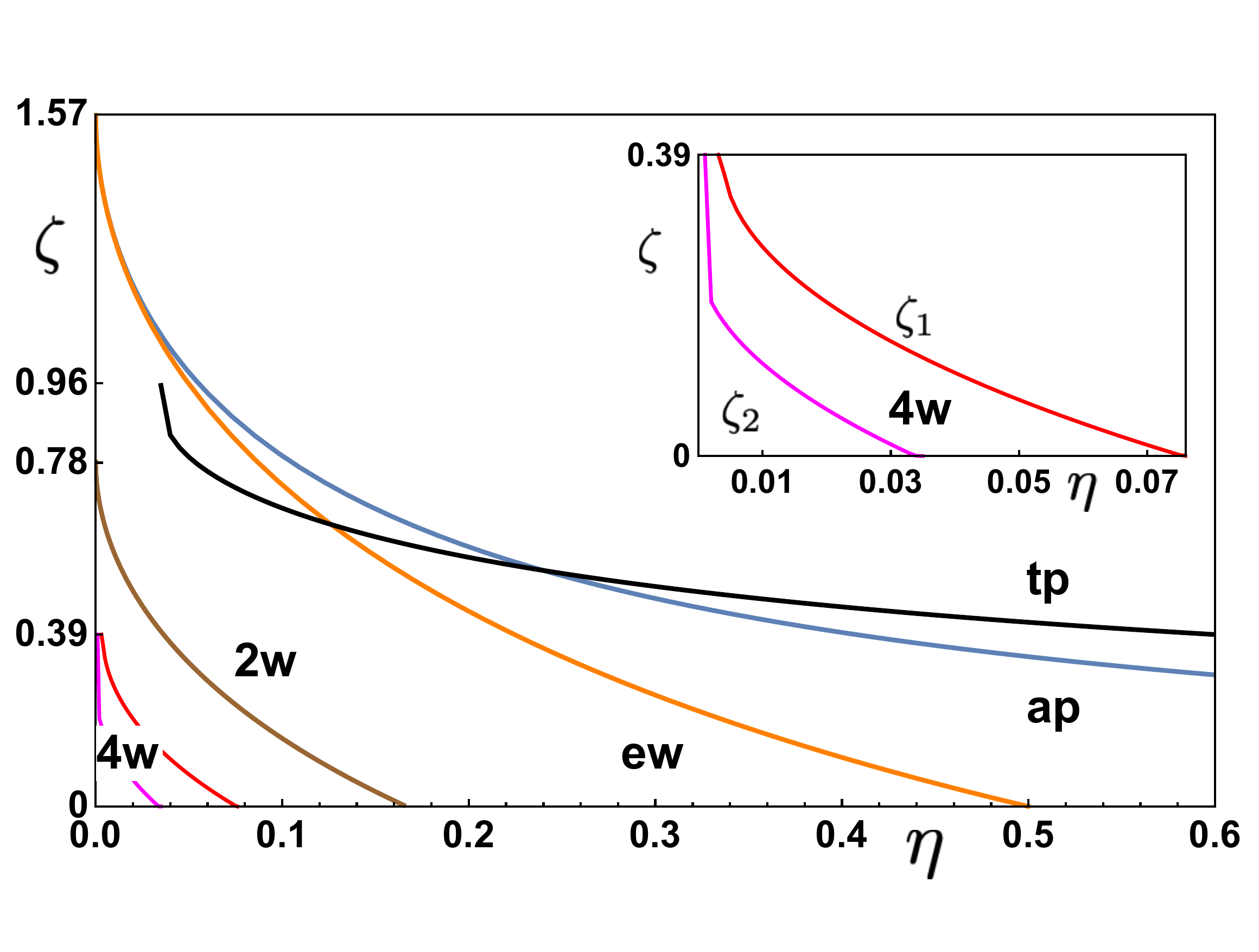} 
\caption{(Color online) 
Optimized core sizes: The symbols ap, and tp represent the antipodal (index 1 each) and tetrahedral (index 1/2 each) configurations of point disclinations; ew, 2w, and 4w refer to the equatorial (index 2), two-wall (index 1 each), and four-wall (index 1/2 each) configurations respectively. The inset depicts the two core sizes required to minimize the total free energy for the 4w-configuration. Order is completely destroyed at $ \zeta = \{ \pi/8, \pi/4, \pi/2, \pi/2 \} $ for the \{4w, 2w, ew, ap\}- configurations respectively. Above $ \zeta = (1/2) \arccos(-1/3) \simeq 0.96 $, cores of neighboring disclinations of the tp-configuration overlap each other (see the text). }
\label{Zeta}
\end{figure}

\begin{figure}[t]
\includegraphics[width=3.2in]{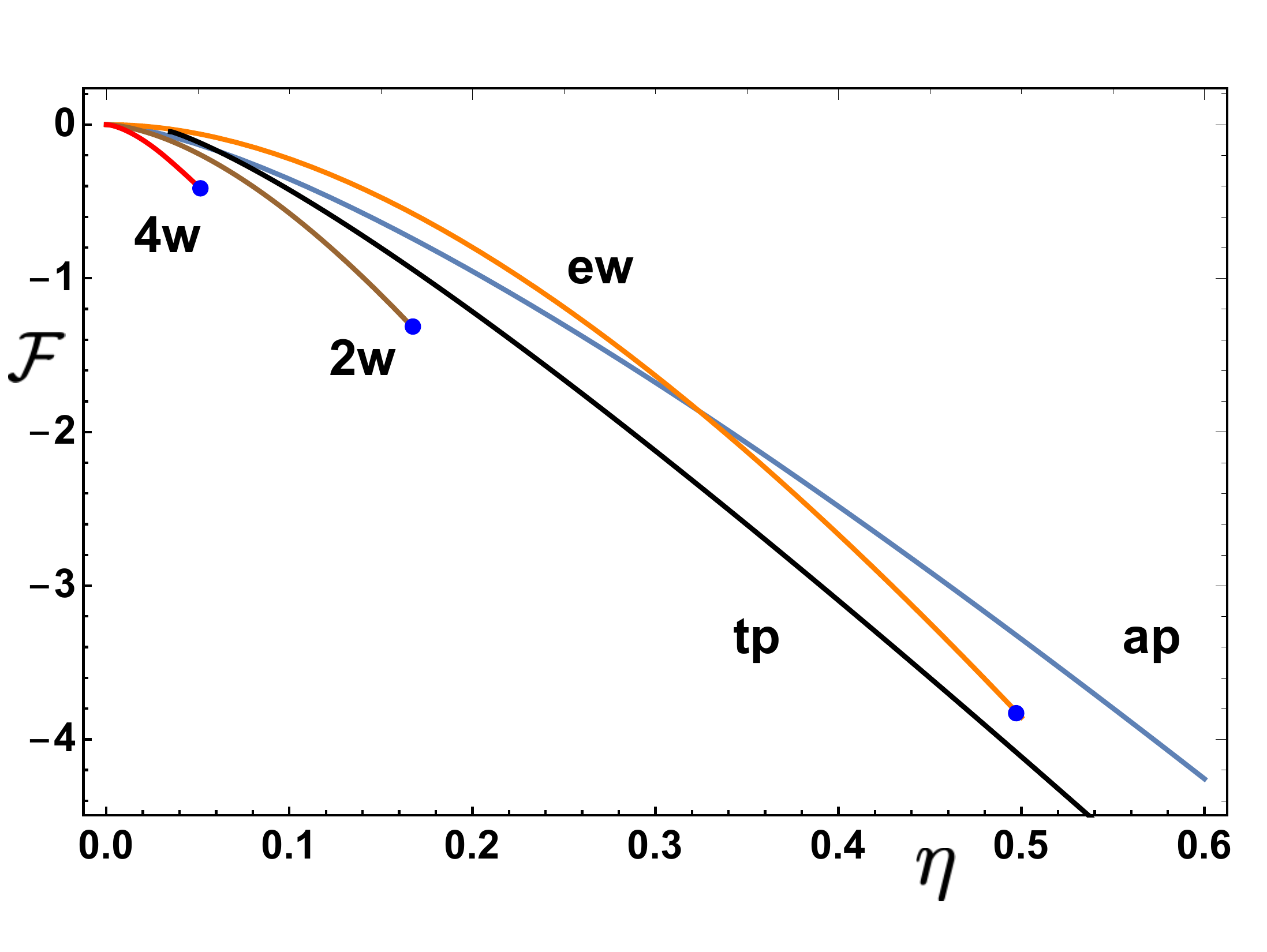} 
\caption{(Color online) 
Minimized free energy $ \mathcal{F} = F_{\mathrm{T}} - 4 \pi \eta $:  We set the reference of  the total free energy  
to  the condensation energy $ 4 \pi \eta $ of the entire sphere.  The dots indicate the $ \eta $-values beyond which wall defects are unstable ($ \zeta(\eta) = 0 $), see Fig.\textcolor{blue}{(\ref{Zeta})}. } 
\label{En}
\end{figure}

Let us consider a unit tangent vector field
$ \hat{\bm{m}}(\theta, \phi) = \cos \alpha(\theta, \phi) \hat{\bm{e}}_{\theta} + \sin \alpha(\theta, \phi)  \hat{\bm{e}}_{\phi} $ 
on a sphere, where we use spherical polar coordinates. For a sphere of radius $ R $, the Gaussian curvature 
$ K = 1/R^{2} $, $ A_{\theta} = 0 $, $ A_{\phi} = - \cos \theta $, and the components of the metric tensor are 
$ g_{\theta \theta} = R^{2}, \, g_{\phi \phi} = R^{2} \sin^{2} \theta, \, g_{\theta \phi} = g_{\phi \theta} = 0 $. 
For the simplest case that describes the antipodal configuration, $ \alpha = 0 $ (streamlines of $  \hat{\bm{m}} $ follow  longitudes on the sphere), with two point disclinations of index 1 each at the poles. The expressions for the optimized core size $ \zeta^{\mathrm{(ap)}} $, and the minimized total free energy  $ \mathcal{F}^{\mathrm{(ap)}} $ of the antipodal configuration are  given in \textcolor{blue}{\cite{SupplMat}}, and are plotted in Fig.\textcolor{blue}{(\ref{Zeta})}, and 
Fig.\textcolor{blue}{(\ref{En})} as functions of $ \eta $. Within the mean field theory, 
$ \eta \sim (T_{c} - T) $. 

We now consider a singular, equatorial wall defect (at $ \theta = \pi/2 $), defined by the disclination density 
\begin{equation}
\label{EqWall}
\mathscr{S}^{(\mathrm{ew})} = \frac{S_{0}}{2 \pi \sqrt{g}} \, \delta (\theta - \frac{\pi}{2}),
\end{equation}
where the superscript $ (\mathrm{ew}) $ stands for equatorial wall. Thus
$ \int_{0}^{2 \pi} \td \phi \int_{0}^{\pi} \mathscr{S}^{(\mathrm{ew})} \sqrt{g} \td \theta = S_{0} $,
where $ S_{0} $ is as yet undetermined. Given this disclination density, we seek a solution to the compatibility condition \textcolor{blue}{(\ref{CompatibilityCond})} subject to boundary conditions. The boundary conditions ensure that there are no point disclinations at the poles (coordinate singularities). This in turn ensures that $ S_{0} = 4 \pi $, vindicating the Poincar\'e-Hopf index theorem \textcolor{blue}{\cite{Brasselet}}. The solution to the compatibility condition is
\begin{equation}
\label{ChiWallSol}
\chi^{(\mathrm{ew})} = 2 \left[ \Theta \left( \theta - \frac{\pi}{2} \right)  \log \cot \frac{\theta}{2} 
                                                              - \log \left( \sqrt{2} \cos \frac{\theta}{2} \right) \right]. 
\end{equation}
The Heaviside $ \Theta $ in  \textcolor{blue}{(\ref{ChiWallSol})} has important consequences for the stability of wall defects, as discussed below.
In terms of $ \alpha $, measured in the local orthronormal frame
$ (\hat{ \bm{e} }_{\theta} (\theta, \phi) , \hat{ \bm{e} } _{\phi} (\theta, \phi)) $, the solution  \textcolor{blue}{(\ref{ChiWallSol})} is particularly simple (Fig.~\textcolor{blue}{\ref{ew}}):  $ \alpha^{(\mathrm{ew})} = - \phi,  \,\mathrm{if} \,  \theta < \pi/2;  \,
\alpha^{(\mathrm{ew})} = \phi, \,\, \mathrm{if} \,\, \pi > \theta > \pi/2 $ \textcolor{blue}{\cite{Footnote2}}.  Substituting for \textcolor{blue}{(\ref{ChiWallSol})} in \textcolor{blue}{(\ref{felChi})}, the elastic free energy of the equatorial wall is 
\begin{equation}
\label{Felew }
F_{\mathrm{el}}^{(\mathrm{ew})}  =   2 \pi K_{\alpha} \{ \sin \zeta - 4 \log \sin [(\zeta/2) + (\pi/4)] - 1 \}.                                              
\end{equation}
We note that the term with the Heaviside $ \Theta $ in \textcolor{blue}{(\ref{ChiWallSol})}, when substituted in $ f_{\mathrm{el}}[\chi] $ \textcolor{blue}{(\ref{felChi})} leads to a term involving $ (\delta [\theta - (\pi/2)])^{2} $ in the integrand. With $ \zeta = 0 $ at, or within the limits of integration, the integral is, strictly speaking, undefined. However it diverges as $1/ \zeta $ \textcolor{blue}{\cite{Footnote3}}, in contrast to the logarithmic divergence encountered in the case of antipodal point disclinations. We therefore expect the index 2 wall to become unstable as $ \zeta $ approaches the molecular size, within the coarse-grained elasticity theory that we have used. This is borne out by the minimization of the total  free energy discussed below.

The condensation free energy is given by  
\begin{equation}
\label{WallCondEn}
F_{\mathrm{C}}^{(\mathrm{ew})} = 4 \pi R^{2} f_{\mathrm{C}} \sin \zeta.
\end{equation}
The core size $ \zeta^{(\mathrm{ew})} $ that minimizes the total free energy $ \mathcal{F}^{\mathrm{(ew)}} = F_{\mathrm{T}}^{(\mathrm{ew})} - 4 \pi \eta $ can be obtained analytically \textcolor{blue}{\cite{SupplMat}}, and is plotted in (Fig.~\textcolor{blue}{\ref{Zeta}}) as a function of $ \eta $. We find that $ \zeta^{(\mathrm{ew})} = 0 $ at $ \eta = 1/2 $. Moreover, it crosses zero and is negative above $ \eta = 1/2 $, which is unphysical. Thus, the singular equatorial wall is stable only for $ 0 < \eta < 1/2 $.  The minimized total free energy of the equatorial wall is (Fig.~\textcolor{blue}{\ref{En}})
\begin{equation}
\label{FTotSpWall}
\mathcal{F}^{\mathrm{(ew)}} = 4 \pi \, [ \log(1 + 2 \eta) - 2 \eta ],
\end{equation}
where we recall that $ \eta = (f_{\mathrm{C}} R^{2})/K_{\alpha} \sim \vert T - T_{c} \vert $. As in the case of antipodal point disclinations, $ \mathcal{F}^{\mathrm{(ew)}} $ depends only on $ \eta $.  

The equatorial wall is not necessarily the minimum energy configuration (ground state for vectorial order) over its entire range of stability. For vector-, as well as nematic order it can split into two walls with index 1 each placed at 
$ \theta = \omega_{1}^{(2 \mathrm{w})} $, and $ \omega_{2}^{(2 \mathrm{w})} = \pi - \omega_{1}^{(2 \mathrm{w})} $, 
where $ \omega_{1}^{(2 \mathrm{w})} < \pi/2 $. In terms of the $ \alpha $ field, the solution to the compatibility condition \textcolor{blue}{(\ref{CompatibilityCond})} for the two-wall configuration is 
$ \alpha^{(2 \mathrm{w})} = ( - \phi,  0,  \phi ) $ in the order of increasing $ \theta $ (Fig.~\textcolor{blue}{\ref{2w}}). Following the same procedure  as used above for the equatorial wall, we find that the numerically minimized total free energy of the two-wall configuration determines the wall locations 
$ \omega_{1}^{(\mathrm{2w})}(\eta) $, and $ \omega_{2}^{(\mathrm{2w})}(\eta) $. The two-wall configuration is stable only for $ 0 < \eta < 0.17 $. 
The angle $ \omega_{1}^{(\mathrm{2w})}(\eta) $ is a monotonic increasing function with $ \omega_{1}^{(\mathrm{2w})}(\eta =0) = \pi/4 $ (corresponding to destruction of order over the entire sphere), and the angle corresponding to zero cutoff can be analytically calculated to be $ \omega_{1}^{(\mathrm{2w})}(\eta = 0.17) = \pi/3 $. At the limit of stability ($ \eta = 0.17 $) the total (integrated) Gaussian curvature of the spherical region between the two walls  $ K_{\mathrm{T} } = 2 \pi $, leaving total Gaussian curvatures of $  \pi $ each for the polar caps. We notice a similar trend for the division of total Gaussian curvature between successive walls in the index 1/2, four-wall configuration for nematic order (see below).

We now discuss the the tetrahedral configuration of point disclinations of index 1/2, followed by the four-wall configuration (of index 1/2 each) for nematic order. It is known \textcolor{blue}{\cite{LubenskyProst}} that for small core sizes the ground state has four disclinations arranged on the vertices of a tetrahedron inscribed in the sphere. For the tetrahedral configuration, we minimize the total free energy numerically, using equal cutoffs along the $ \theta $- and $ \phi $ directions to obtain the optimized core size (Fig.~\textcolor{blue}{\ref{Zeta}}), and the total free energy (Fig.~\textcolor{blue}{\ref{En}}).  Above $ \zeta \simeq 0.96 $ (below $ \eta(\zeta = 0.96) \simeq 0.05 $), cores of neighboring disclinations overlap each other at nonzero $ \eta $, and an extension of the approach of \textcolor{blue}{\cite{ParkLubMcK}} is better suited to address the problem. In this work, we do not pursue this approach.

The four-wall configuration (Fig.~\textcolor{blue}{\ref{4w}}) has two walls located at 
$ \theta = (\omega_{1}^{(\mathrm{4w})}, \, \omega_{2}^{(\mathrm{4w})} > \omega_{1}^{(\mathrm{4w})}) $ on the upper hemisphere. The other two symmetry related walls are located in the lower hemisphere, with the convention 
$ \omega_{3}^{(\mathrm{4w})} < \omega_{4}^{(\mathrm{4w})} $. The solution to the compatibility condition gives  $ \alpha^{(4 \mathrm{w})} = ( - \phi, \, - \phi/2,\,  0, \, \phi/2, \, \phi ) $, in the five regions arranged in the order of increasing $ \theta $. We find that this configuration is stable only for $ 0 < \eta <  0.05 $.  In minimizing the total free energy of the pairs of walls we need to use two core sizes, $ \zeta_{1}^{(\mathrm{4w})} $ and $ \zeta_{2}^{(\mathrm{4w})} $, for the walls at  $ \omega_{1}^{(\mathrm{4w})} $ and $ \omega_{2}^{(\mathrm{4w})} $ respectively. The angles $ \omega_{1}^{(\mathrm{4w})} $ and $ \omega_{2}^{(\mathrm{4w})} $ are weakly monotonic increasing functions of $ \eta $, with $  \omega_{1}^{(\mathrm{4w})}(\eta = 0.05) = 2 \arctan (1/\sqrt{7}) (\simeq 0.72) $, and  $  \omega_{2}^{(\mathrm{4w})}(\eta = 0.05) = 2 \arctan (\sqrt{(3/5}) (\simeq 1.32) $. We 
 notice that putting the exact values as limits of integration, we get $ K_\mathrm{T} = \pi $, indicating a systematic division of total Gaussian curvature between neighboring walls. In what follows we trace this trend to the elastic part of the free energy, and generalize it to $ n $-atic order.

We find that configurations involving intersecting walls (for example, longitudinal walls along $ \phi = 0 $ and $ \phi = \pi/2 $, vectorial order) have higher free energies  than the configurations discussed above. This is because of the larger repulsive interaction energy near their intersection points, where the walls are close. 

Wall defects are stable in the range $ 0 < \eta < 0.5 $. We use mean field theory to examine the extent to which wall defects are realizable in experiments. This involves estimating the radius $ R $, the thickness $ h $, and the temperature range 
$ \Delta T = T_{c} - T $ over which stable wall defects are observable. We recall that the scale of $ \eta $ in Figs. 4-5 is linear in $ \Delta T $. 

For the sake of concreteness we consider the experiments of \textcolor{blue}{ \cite{DeVries}}, where molecules of the self assembled monolayer (thickness $ h $ of order molecular length) on the sphere are tilted with respect to the local normal to the sphere. The projection of tilted molecules on to the local tangent plane to the sphere imparts vectorial order to it. 
Within mean field theory the correlation length diverges as $ \xi = \xi_{0} \, (\Delta T/T_{c} )^{-1/2} $. The bare correlation length $ \xi_{0}  $ is of order $ 2 \mathrm{nm} $ (molecular dimensions), and $ K_{\alpha} \simeq k_{\mathrm{B}} \Delta T $, leading to 
$ f_{\mathrm{C}} \simeq  k_{\mathrm{B}} \Delta T/ \xi^{2} $. Substituting for $ \xi $ in $ \eta $, we get 
$ \Delta T \simeq  T_{c} (\xi_{0} /R)^{2} \eta $. For $ T_{c} \simeq 300 \mathrm{K} $,  and spheres with $ R =  5 \mathrm{nm} $, and $ R =10 \mathrm{nm} $ 
(used in \textcolor{blue}{\cite{DeVries}}), $ \eta = 0.1 $ corresponds to $ \Delta T \simeq 4.8 \mathrm{K} $, and $ \Delta T \simeq 1.2 \mathrm{K} $ respectively, thus establishing the temperature scale.  However, these sizes may be too small for observing wall defects. For $ R \simeq 35 \mathrm{nm} $, $ \eta \simeq 0.1 $ corresponds to $ \Delta T \simeq 0.1 \mathrm{K} $. Nanoparticles with radius around $ 35 \mathrm{nm} $ are likely to be good candidates for observing wall defects.
For $ R = 35 \mathrm{nm} $ the equatorial wall is stable between $ \eta \simeq 0.16 \, (\equiv \Delta  T = 0.16 \mathrm{K}) $ and $ \eta = 0.5 \, (\equiv \Delta T = 0.5 \mathrm{K}) $; it is likely to be the simplest one to observe. Evidently, for large core sizes, point- as well as wall defects will not have a ``valence" = 1, as is  the case for the antipodal disclinations of \textcolor{blue}{ \cite{DeVries}}. 

We now generalize, and make precise our remarks on the systematics of the division of total Gaussian curvature between neigboring walls. For core sizes $ \zeta \rightarrow 0 $ the condensation energy is negligibly small, clearly implicating the elastic free energy as the root cause of this phenomenon.  By minimizing the \textit{elastic} free energy, we find that for $ n $-atic order on spheres the lowest elastic free energy configuration has $ 2 n $ walls of index $ 1/n $ each, located such that the integrated Gaussian curvature between any two successive walls
\begin{equation}
\label{KtotalDivision}
K_{\mathrm{T}}(i, i + 1) = 2 \pi \int_{\omega_{i}}^{\omega_{i + 1}} K \sqrt{g} \td \theta = 2 \pi/ n,  
\end{equation}
where $ i = \{1, 2,..., 2 n -1 \} $ labels the walls in the order of increasing $ \theta $. The details of the calculations leading to \textcolor{blue}{(\ref{KtotalDivision})} are given in \textcolor{blue}{\cite{SupplMat}}.  

In this letter we have used a simplified version of the Ginzburg-Landau theory to predict the existence of stable topological wall defects (near the order-disorder transition) on spheres with $ n $-atic order. Experiments and simulations can test our results. In our analysis we have ignored the effects of thermal fluctuations. Fluctuation effects will be important close to $ T_{c} $. However, spheres are closed surfaces, and the system size is very small, thus diminishing the effects of fluctuations. The nature of order-disorder transition on orientationally ordered spheres is not clear, and needs to be investigated. In particular, the transition may not be of the Kosterlitz-Thouless type (see, e.g. \textcolor{blue}{\cite{ChaikinLubensky}}). \textit{Close to $ T_{c} $, interacting wall defects, rather than point defects will dominate the transition.}  A detailed analysis of fluctuation effects is beyond the scope of this paper. In addition to fluctuation effects, it  would be of interest to extend the theory to include the effects of anisotropy of elastic constants, and study the shape-changes of deformable vesicles, induced by wall defects. We have investigated the stability of topological wall defects on catenoids \textcolor{blue}{\cite{SaichandEtAl}}, which will be published elsewhere.

We thank T. C. Lubensky and N. V. Madhusudana for useful discussions, and Md. Arsalan Ashraf for help with graphics. Thanks are due to V. A. Raghunathan for many instructive discussions, and to Joseph Samuel for helpful suggestions and a careful reading of the manuscript.

\newpage
\pagebreak

\onecolumngrid
\appendix

\section{SUPPLEMENTAL MATERIAL}

\section{Energetics of antipodal disclinations:}

The elastic free energy of the vectorial texture outside the core region is
\begin{equation}
\label{SphAntipodal}
 F_{\mathrm{el}}^{\mathrm{(ap)}} = 2 \pi K_{\alpha} \, [ \log \cot (\zeta/2)  - \cos\zeta ],
\end{equation}
where the superscript $ (\mathrm{ap}) $ stands for antipodal point disclinations, and the dimensionless cutoff $ \zeta =  r_{\mathrm{c}}/R $.  For small $ r_{\mathrm{c}} $ this result reduces to that of reference \textcolor{blue}{ [2] }  of the main text. 
We note that $ F_{\mathrm{el}}^{\mathrm{(ap)}} $  depends solely on $ \zeta $, and diverges logarithmically as $ \zeta \rightarrow 0 $. The condensation free energy is 
\begin{equation}
\label{fC}
F_{\mathrm{C}}^{\mathrm{(ap)}} = 4 \pi R ^{2} f_{\mathrm{C}} \, (1 - \cos \zeta ). 
\end{equation}

For antipodal points the optimum core size is 
\begin{equation}
\label{CoreAP}
\zeta^{(\mathrm{ap})} = 2 \arctan[p(\eta)/q(\eta)],
\end{equation}
where 
$ p(\eta) = [1 + 2 \eta - \sqrt{2 \eta (1 + 2 \eta)}  \,]^{1/2}  $, and $ q(\eta) =  [1 + 2 \eta + \sqrt{2 \eta (1 + 2 \eta)} \,]^{1/2 } $. 
\\  

The minimized total (elastic + condensation), dimensionless free energy of the antipodal configuration in units of $ K_{\alpha} $ is 
\begin{equation}
\label{Fap}
\mathcal{F}^{\mathrm{(ap)}} =  2 \pi [ 2 \eta - \sqrt{2} \, h(\eta) + \arctanh (\sqrt{2} \, \eta/h(\eta))] - 4 \pi \eta,
\end{equation}
where $ h(\eta) = \sqrt{\eta (1 + 2 \eta)} $, and we have taken the dimensionless condensation free energy $ 4 \pi \eta $ for destruction of order over the entire sphere as the reference. \\

\section{Solution of the compatibility condition for the equatorial wall:}

The general solution to the compatibility condition (equation \textcolor{blue}{(3)} of the main text) is
\begin{equation}
\label{ChiGenSol}
\chi(\theta) = - a_{1} \log \tan(\theta/2) - \log \sin \theta  
                       + [S_{0}/(2 \pi)] [ \Theta(\theta - (\pi/2)) -1] \log \cot (\theta/2) + a_{2},
\end{equation}                              
where the symbol $ \Theta $ is the Heaviside theta, and $ a_{1}, \, a_{2} $ are constants. Setting 
$ a_{2} = 0 $, we exploit the symmetry $ \chi(\theta) = \chi(\pi - \theta) $ to obtain $ a_{1} = - S_{0}/(4 \pi) $.
Note that $ S_{0} $ is as yet undetermined. To ensure that there are no point disclinations of index 1 each at the north and south poles, we investigate the behavior of $ \partial_{\theta} \, \chi(\theta) $ at the poles. We note that $ \lim_{\theta \to 0} \partial_{\theta} \, \chi(\theta) $ as well as  $ \lim_{\theta \to \pi} \partial_{\theta} \, \chi $ go to infinity unless
$ S_{0} = 4 \pi  $. With $ S_{0} = 4 \pi $, both these limits go to zero.  Setting $ a_{1} = - S_{0}/(4 \pi)= -1 $ guarantees that there are no point disclinations at the north and south poles, and yields the solution, equation \textcolor{blue}{(6)}  of the main text. We have thus constructed a wall defect with index 2, as demanded by the Poincar\'e-Hopf theorem extended to non-isolated zeros.  \\

\section{Energetics of the equatorial wall:}

The optimum core-size of the equatorial wall defect is  
\begin{equation} 
\label{CoreSpWall}
\zeta^{(\mathrm{ew})} = \arctan \frac{ 1 - 2 \eta }{ 2 \sqrt{2 \eta } }.
\end{equation}  

The minimized, dimensionless total free energy of the equatorial wall in units of $ K_{\alpha} $, with reference to  the dimensionless condensation free energy $ 4 \pi \eta $ is
\begin{equation}
\label{Few}
\mathcal{F}^{(\mathrm{ew})} =  4 \pi  \left[  \log \left( \cot \frac{\pi +   \zeta^{(\mathrm{ew})}}{4} \right)  -  \log \cos \zeta^{(\mathrm{ew})}  +  (1 + \eta) \sin \zeta^{(\mathrm{ew})} + (\log 2 - \frac{1}{2})  \right] - 4 \pi \eta.
\end{equation} 

Further simplification of \textcolor{blue}{(\ref{Few})} above leads to  equation \textcolor{blue}{(9)} of the main text.  \\

\section{Division of Gaussian curvature by wall defects:}

We consider $ n $-atic order on a sphere with $ 2 n $ walls of strength $ 1/n $ each, and indicate the positions of symmetry related pairs of walls by $ (\omega_{k}, \omega_{-k} = \pi - \omega_{k}) $, where 
$ k = 1, 2,..., n $. Our aim is to minimize the elastic free energy of such configurations with respect to the angular positions of the symmetry-related pairs of walls. This leads to the result on the division of Gaussian curvature presented in the main text. \\

The disclination density of this configuration is
\begin{equation}
\label{WallDisclDen}
\mathscr{S}(\theta) = \frac{1}{n \sqrt{g}} \, \sum_{k = 1}^{n} \, [\delta(\theta - \omega_{k}) + \delta(\theta - \omega_{- k})].
\end{equation} 
To evaluate the elastic free energy of the configuration we use the Coulomb gas form of the elastic free energy. The elastic energy $ F_{\mathrm{el}} $ of the main text can be written as (see reference \textcolor{blue}{ [7] } of the main text)
\begin{equation}
\label{FelCoulomb}
F_{\mathrm{el}} = - \, \frac{1}{2} K_{\alpha} \iint \rho(\theta, \phi) \, 
                                 G(\theta, \phi; \, \theta', \phi') \, 
                                \rho(\theta', \phi') \td \mathcal{A} \td \mathcal{A}',
\end{equation}
where $ \rho(\theta, \phi) = K( \theta, \phi) - \mathscr{S}(\theta, \phi) $. The Green's function $ G(\underline{\bm{\sigma}}, \underline{\bm{\sigma}}') $ for  a spherical surface  is 
\begin{equation}
\label{GreenFnSph}
 G(\theta, \phi; \, \theta', \phi') = \frac{1}{4 \pi} \log \left[ \frac{1 - C_{\beta}(\theta, \phi; \, \theta', \phi')}{2} \right],
\end{equation}
where 
\begin{equation}
\label{Cbeta}
C_{\beta}(\theta, \phi; \, \theta', \phi') = \cos \theta \, \cos \theta' + \sin \theta \, \sin \theta' \, \cos( \phi - \phi').
\end{equation}
Given the azimuthal symmetry of the configuration (no $ \phi $-dependence), the Green's function satisfies the Laplace equation
\begin{equation}
\label{GreenLaplace}
\sin \theta \, \partial_{\theta}^{2} G(\theta; \theta') + \cos \theta \, \partial_{\theta} G(\theta; \theta') 
                                          + \frac{ \sin \theta }{ 4 \pi } = \frac{ \delta(\theta, \theta') } { 2 \pi },
\end{equation}                                          
with the conditions $ G(\theta; \theta') = G(\theta'; \theta) $, and $ G(\pi - \theta; \pi - \theta') = G(\theta; \theta') $. The solution of the above equation is
\begin{equation}
\label{GreenSol}
G(\theta; \theta') = \frac{1}{8 \pi} [ \log (\sin \theta \, \sin \theta') 
  +  ( \Theta(\theta' -\theta) - \Theta(\theta -\theta') ) \, \log( \cot \frac{\theta}{2} \, \tan \frac{\theta'}{2} )  ], 
\end{equation}
where the Heaviside theta is defined as $ \Theta(x) = 1 $ if $ x > 0 $, and $ \Theta(x) = 0 $ if $ x < 0 $.  \\

The elastic free energy  \textcolor{blue}{(\ref{FelCoulomb})}  can be written as 
$  F_{\mathrm{el}} = - (1/2) K_{\alpha} ( F_{KK}   -  2 F_{K \mathscr{S}} + F_{\mathscr{S S} } ) $, where  
\begin{align}
\label{FCoulombExp}
F_{KK}                       &=  \iint K  \, G(\theta; \, \theta') 
                                           \, K \td \mathcal{A} \td \mathcal{A}',         \nonumber   \\                                                                                  
F_{K \mathscr{S}}      &=   \iint K \, G(\theta; \, \theta') \, 
                                            \mathscr{S} (\theta') \td \mathcal{A} \td \mathcal{A}', \,  \mathrm {and} \nonumber  \\
 F_{\mathscr{S S} }    &=    \iint \mathscr{S}(\theta) \, G(\theta; \, \theta') \, 
                                            \mathscr{S}(\theta') \td \mathcal{A} \td \mathcal{A}'. 
\end{align} 
Substituting for $  G(\theta; \theta') $ and for  $ \mathscr{S}(\theta) $, and minimizing with respect to the angular positions $ \omega_{i} $, we find that only the $ F_{\mathscr{S S} } $ term is important in determining $ \omega_{i} $.  After some algebraic manipulations, we find that the set of angular positions that minimizes the elastic free energy is given by
\begin{equation}
\label{omegaiMin}
2 \cos \omega_{i} = \frac{ 2 n - 2 i +1}{n}, 
\end{equation} 
and that the integrated Gaussian curvature between the symmetry-related walls at $ \omega_{i} $ and $ (\pi - \omega_{i}) $ is 
$ 2 \pi (2 \cos \omega_{i})  = \frac{2 \pi}{n} (2 n - 2 i +1) $. This directly leads to the result, equation \textcolor{blue}{(10)} of the main text.


\begin{thebibliography}{9}
\bibitem{Footnote1} $ \mathscr{S}(\underline{\bm{\sigma}}) = (2 \pi/\sqrt{g}) \, \sum_{m} q_{m} \, 
                                                                   \delta^{(2)}(\underline{\bm{\sigma}} - \tilde{\underline{\bm{\sigma}}}_{m}) $  for isolated disclinations with point disclination charges $ q_{m} $ located at  $ \tilde{\underline{\bm{\sigma}}}_{m} $.
The charges $ q_{m} $ have discrete values allowed by the $ n $-atic orientational order.
\bibitem{Footnote2} Notice that 
$ \oint_{0}^{2 \pi} \td \alpha_{\mathrm{(lh)}}  - \oint_{0}^{2 \pi} \td \alpha_{\mathrm{(uh)}} = 4 \pi $, where 
$ \alpha_{\mathrm{(lh)}} $ and $ \alpha_{\mathrm{(uh)}} $ refer to the lower and the upper hemisphere respectively, with the integrals taken in the anticlockwise sense around the outward normal at the respective poles. Thus the index $ 4 \pi/(2 \pi) = 2 $ for the equatorial wall. This result is easily generalised to walls having other indices.
\bibitem{Footnote3} As can be seen by putting $ \theta = 0 $ in $ \delta(\theta) = (1/(2 \pi))\int_{-\infty}^{\infty} e^{i k \theta} \td k $, and setting the ultraviolet cutoff to $ 2 \pi/\zeta $.
\end{thebibliography}
\end{document}